\documentclass[aps,prl,twocolumn,groupedaddress,showpacs]{revtex4}
\usepackage{graphicx}
\graphicspath{{figure file fold path}}
\usepackage[dvipsnames,usenames]{color}
\usepackage{color}
\usepackage{amsmath}
\usepackage{float}
\usepackage{amssymb}
\usepackage{amsthm}
\usepackage{mathrsfs}

\emergencystretch=\maxdimen
\begin{document}

\title{Fast square-oscillations in semiconductor VCSELs with delayed orthogonal polarization feedback}
\author{Tao Wang$^{1}$, Zhicong Tu$^{1}$, Yixing Ma$^{1}$, Yiheng Li$^{1}$, Zhibo Li$^{1}$, Fan Qin$^{1}$, Stephane Baland$^{2}$, and Shuiying Xiang$^{1}$}

\affiliation{$^1$State Key Laboratory of Integrated Service Networks, Xidian University, Xi'an 710071, China}
\affiliation{$^2$Universit\'e C\^ ote d’Azur, Institute de Physique de Nice, UMR 7010 CNRS, Nice, 06200, France}

\date{\today}

\begin{abstract}
We present an experimental investigation into the generation of self-sustained and fast square oscillations from the TE mode of semiconductor VCSELs with delayed orthogonal polarization feedback. We find that the low frequency switching originates from the rotation of the TE and TM modes facilitated by a long time delay, but the fast oscillations are anchored to the frequency beating between the TE and TM modes and are modified by a half-wavelength ($\lambda/2$) plate. A comprehensive analysis of the evolution of the nonlinear dynamics is conducted and the related mechanism is discussed. Our study not only deepens our comprehension of laser nonlinear dynamics but also offers an all-optical approach for producing specialized signals, which could be instrumental in applications such as optical communications and photonic computing leveraging the complexity of long-delay systems. 
\end{abstract}


\maketitle 

\section{Introduction}
Semiconductor lasers, especially Vertical Cavity Surface Emitting Lasers (VCSELs), subject to external-cavity optical feedback are considered as ideal systems for the study of nonlinear dynamics. The interaction among different frequency oscillations, such as relaxation oscillations, mode-beating frequencies, polarization switching rates, external-cavity frequencies, and current modulation frequencies, can give rise to rich dynamics including chaos, periodic oscillations, and regular pulse trains can be generated~\cite{Tiana-Alsina2022, Wang2023, Skenderas2024}. Those dynamical behaviors possess various potential applications in secure optical communications~\cite{Wu2013, Xue2016}, high speed random bit generators~\cite{Argyris2016, Li2016OL}, chaotic radar~\cite{Myneni2001, Lin2004}, and photonic spiking neural networks~\cite{Xiang2024, Inagaki2021}. Thus, the investigation on the nonlinear system of semiconductor lasers with external optical feedback has attracted growing interest and research attention, not merely from a fundamental physics perspective, but also for their new potentials on enabling practical applications.

Among various dynamical regimes, square-wave (SW) polarization switching represents a critical operational state with distinct technological implications. SW oscillations provide essential clock signals for optical time-division multiplexing~\cite{Li2016} and phase-encoded communications~\cite{Mossaad2024}. In edge-emitting lasers (EELs), crossed-polarization re-injection (XPR) schemes have demonstrated SW generation via asymmetric TE (transverse electric) and TM
(transverse magnetic) modes coupling, where the oscillation period ($T_{SW} \approx 2\tau$) arises from delayed polarization bistability~\cite{Gavrielides2006, Gavrielides2010}. Subsequent studies in polarization-coupled EEL pairs revealed subharmonic synchronization phenomena with ($T_{SW} \approx \tau$) and tunable duty cycles~\cite{sukow2012square,Masoller2011}, highlighting the critical role of delayed cross-polarization coupling. Then, the SW generation paradigm extends uniquely to VCSELs due to their inherent polarization properties. Unlike EELs, VCSELs exhibit strong linear polarization selection originating from crystalline anisotropy and asymmetric strain~\cite{Yu2009, Yariv2007}, with orthogonal polarization mode suppression exceeding 30 dB in optimized designs~\cite{Pan2024}. Polarization-rotated optical feedback (PROF) induces counterintuitive dynamics: despite the dominant TE mode lasing, delayed TM-mode reinjection triggers deterministic polarization switching through nonlinear mode competition~\cite{Islam2021, Mulet2007, Marconi2013}. 

Analyzing SW fast dynamics is also crucial for understanding transient polarization switching, coherence collapse, and nonlinear mode competition, which are essential for stabilizing high-speed optical waveforms. Studies by Virte et al. has explored deterministic polarization chaos and square-wave-like switching in VCSELs, linking fast dynamics to nonlinear mode competition~\cite{virte2013deterministic}. Additionally,  Sukow et al. have demonstrated SW self-oscillations in VCSELs under feedback, and analyzed the evolution of the switching~\cite{sukow2012square}. The importance of examining square-wave fast oscillations lies in their potential applications in ultrafast optical communication and neuromorphic computing, where rapid and deterministic polarization switching enables high-bit-rate signal encoding and low-power, noise-resistant processing.


In this work, we present experimental observations of square-wave (SW) polarization switching and ultrafast oscillations in a semiconductor VCSEL subjected to delayed orthogonal-polarization feedback. The VCSEL operates in two linear polarization modes -- TE and TM modes -- with a significant threshold disparity. Within the investigation region, the TE mode dominates the lasing emission, while the TM mode is significantly suppressed. By employing an external ring cavity with a $\lambda/2$ wave plate to rotate the polarization of both modes, we comprehensively investigate SW dynamics as functions of pump current and feedback phase. Our findings reveal that both the amplitude and duration of the SW can be adjusted by modifying these parameters. We also focus on the fast polarization dynamics which accompanies the SW and show that the non-zero part of the SW is in fact dominated by very fast oscillations which in general maintain very well defined phase relation from one round-trip to the next. Our experiment is realized in a long-delay regime (\textit{i.e.} the reinjection occurs after a long time with respect to typical semiconductor laser time scales). This confers it a large number of degrees of freedom whose space-like dynamics has been discussed in \cite{marino2014front,garbin2015topological,marconi2015vectorial,romeira2016regenerative} and subsequent works. Thus, the platform constitutes a novel and practical testbed for the analysis of space-like phenomena in long-delayed systems, whose complexity may be helpful for neuromorphic photonic computing~\cite{Friart2014, Uy2018}.

\section{Experimental setup}
\begin{figure}[ht!]
\centering
  \includegraphics[width=7.5cm]{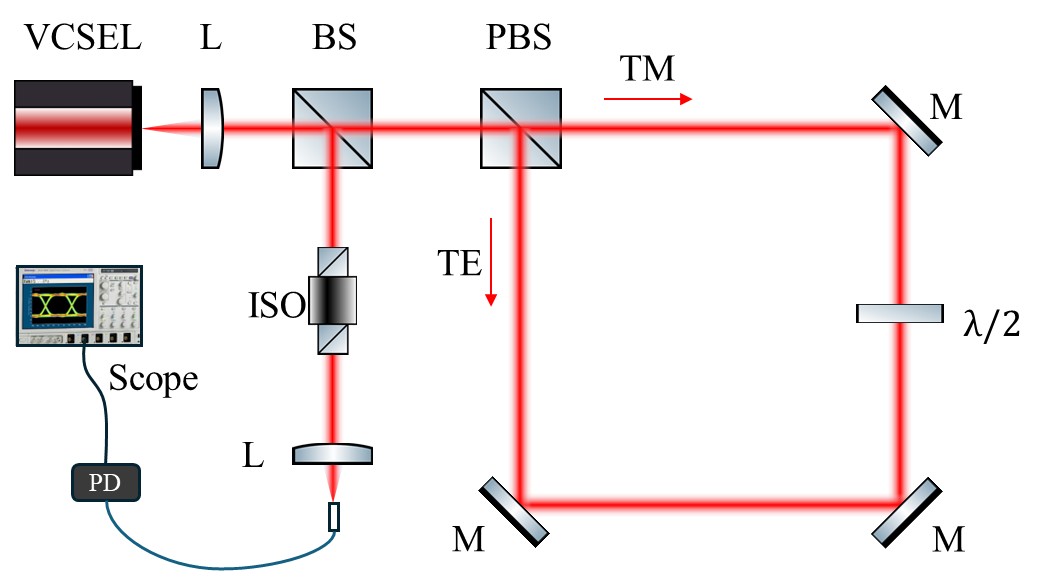}
  \caption{Experimental setup: L, optical lens; BS, beam splitter; PBS, polarized beam splitter; M, mirror; $\lambda/2$, half-wave plate; ISO, optical isolator; PD, amplified high-speed photodetector; Osc., oscilloscope.}
  \label{Setup}
\end{figure}

Fig.~\ref{Setup} depicts the schematic diagram of the experimental design, where a temperature-stabilized semiconductor VCSEL (Thorlabs L850VH1, $\lambda = 850$ nm at $T=25^\circ$C) is employed as the light source. The laser is operated in single longitudinal and transverse modes, and is driven by a low-noise current source (Thorlabs LDC205C). The temperature is stabilized at 25$^\circ$C by a temperature controller (Thorlabs TED200C), better than 0.001$^\circ$C. The collimated laser light initially passes through a beam splitter (BS, 50:50), causing the beam to be divided into two equal-intensity beams. One beam goes to an amplified fast detector (10 GHz bandwidth) after passing through an optical isolator, and the temporal signals are monitored by a 20 GHz oscilloscope (Tektronix DSA72004, sampling rate is 50 GHz). Meanwhile, the second beam is directed towards an external ring cavity. Upon passing through a polarized beam splitter (PBS), the TE and TM modes are separated and circulate in opposing directions within the external cavity. After traversing a $\lambda/2$ plate, they undergo a 90$^\circ$ rotation. The total feedback path length is 1.8 m, which corresponds to a round-trip time of 12 ns. The optical isolator’s input polarizer is aligned to selectively monitor the dynamics of the TE mode.

\section{Results and discussions}
\begin{figure}[ht!]
\centering
  \includegraphics[width=8.5cm]{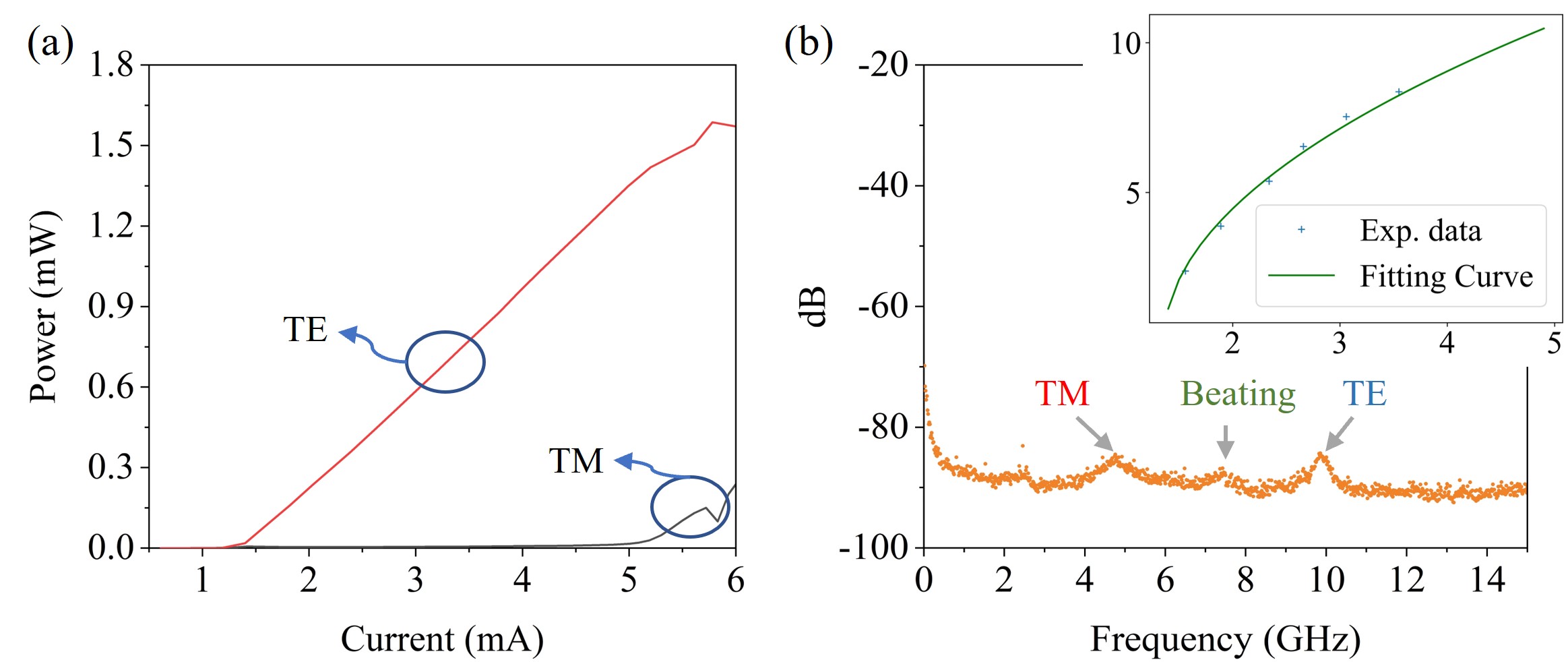}
  \caption{Lasing function curves and typical RF spectrum of the free running VCSEL: (a) input-output function curves as a function of pump current. The red curve indicates the TE mode, and the black curve denotes the TM mode; (b) typical RF spectrum obtained from the electrical spectrum analyzer (ESA) characterization after balancing the polarization. Inset, relaxation oscillation frequency of the free running laser's TE mode as a function of the pump current. }
  \label{lasercurvespectrum}
\end{figure}

Fig.~\ref{lasercurvespectrum}a presents polarization-resolved light-current (L-I) characteristics of the VCSEL, revealing distinct threshold behaviors: the TE mode (red curve) exhibits a significantly lower threshold than the TM mode (black curve), specifically $P_{th}^{TE} =$ 1.48 mA and $P_{th}^{TM} =$ 5.20 mA $=3.51P_{th}^{TE}$ . Within the range of $P_{th}^{TE} < P < P_{th}^{TM}$, the output power of the TE mode increases linearly with the input current. However, the TM mode is strongly suppressed, and its signal is approximately close to zero. When $P > P_{th}^{TM}$, the output power of the TM mode starts to increase. In this region, fluctuations at low frequency are observed in the RF spectrum of both the TM and TE modes, suggesting complex polarization dynamics taking place. Fig.~\ref{lasercurvespectrum}b shows the radio frequency (RF) spectrum obtained from the free running laser at $P = 5.62~mA = 3.8 P_{th}^{TE}$. For specific orientation of the isolator's input polarizer, we observe three distinct peaks located at 4.8 GHz, 7.5 GHz, and 9.8 GHz, respectively. The inset shows the relaxation oscillation frequency of the TE mode as a function of current, along with the corresponding square root-fit. Upon a detailed examination of the curve for the frequency at $P =$ 3.8$P_{th}^{TE}$, we attribute the peak at 9.8 GHz to the relaxation oscillation frequency of the TE mode and the peak at 4.8 GHz to the relaxation oscillations of the TM mode. The central peak at 7.5~GHz can be interpreted as the beating note between the TE and TM modes and therefore corresponds to the VCSEL's birefringence (in agreement with typical birefringence order of magnitude of 5~GHz\cite{Mulet2007}).

\begin{figure}[ht!]
\centering
  \includegraphics[width=8.5cm]{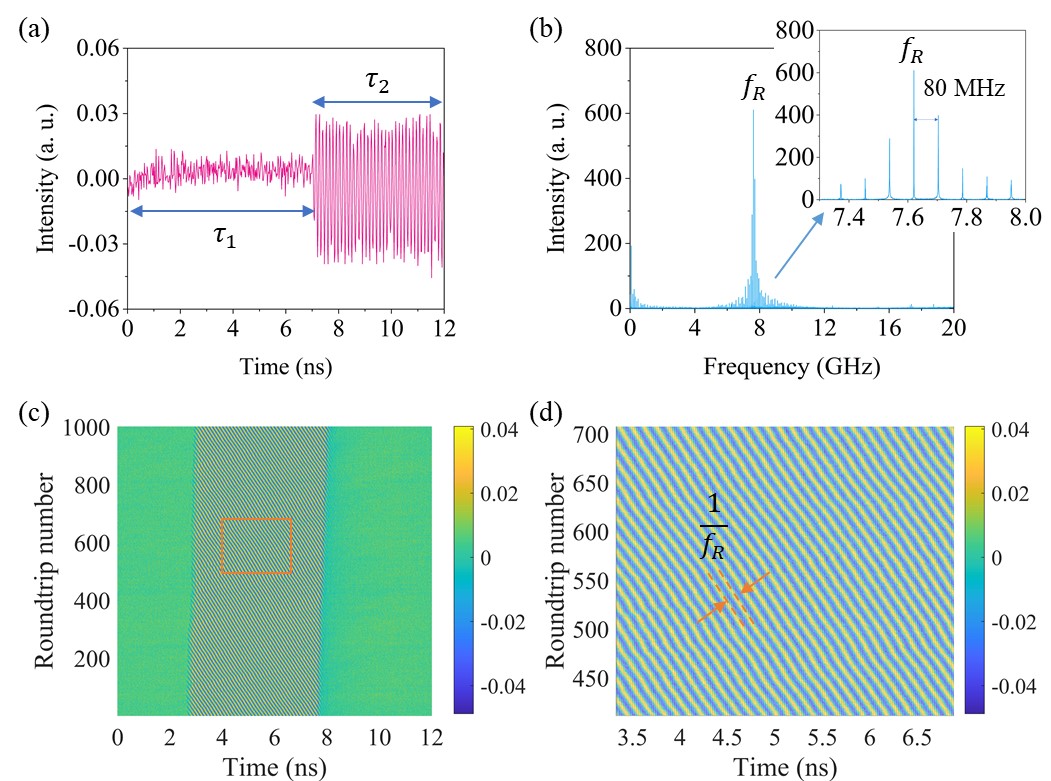}
  \caption{Square wave characterization: (a) temporal dynamics within one period; (b) RF spectrum corresponding to the temporal signal. Inset, enlarged central frequency region; (c) spatiotemporal diagram of the square wave dynamics when the pump current is 4.24 mA and the $\lambda/2$ is 96$^\circ$; (d) the highlighted structure, which is outlined by the orange square in (c).}
  \label{squareDynamics}
\end{figure}

Fig.~\ref{squareDynamics}a illustrates the temporal waveform within a single period $\tau = \tau_1 + \tau_2$, corresponding to one round-trip. The waveform exhibits two distinct alternating temporal states: one is the steady state accompanied by background noise, lasting $\tau_1 = 7.0$ ns, while the other one is characterized by fast oscillations, with a duration of $\tau_2 = 5.0$ ns. It is important to note that the noise waveform is attributed to the detection noise. Therefore, given that the TM mode operates below the threshold with nearly zero intensity, the switching occurs between a steady state for the TM mode and fast oscillations associated with the TE mode. 

Fig.~\ref{squareDynamics}b displays the corresponding RF spectrum obtained after performing a Fourier transform on the temporal signal sequence. It is clear that a prominent central peak is located at $f_R = 7.6$ GHz, signifying the rapid oscillation frequency of the TE mode, which is consistent with the birefringence measured in the free-running laser. Moreover, a series of side modes are observed, which contribute to a comb-like structure in the spectrum. The side modes originate from the longitudinal modes of the external cavity, indicating complex interactions that stem from the feedback loop dynamics. Through a closer inspection, it is found that the spacing between the side modes is consistently measured at 80 MHz. This specific spacing can be interpreted as being equivalent to twice the feedback delay. The phenomenon suggests that, after passing through the $\lambda$/2-plate once, the TE mode undergoes conversion into the TM mode and injected into the VCSEL in that state. Due to the birefringence of the VCSEL, a substantial portion of this injected light reflected back into the feedback loop. Subsequently, the light is converted back to the TE mode and efficiently couples into the laser. Such dual-pass polarization conversion seems to be ubiquitous in polarization-rotated feedback experiments~\cite{oliver2011dynamics}.

Fig.~\ref{squareDynamics}c shows the space-time diagram of the typical square-oscillations from the TE mode when the laser is operated at 2.86$P_{th}^{TE}$ and the angle of the $\lambda/2$-plate is set at 96$^\circ$. The construction of this diagram is based on an analogy between time delayed and spatially extended systems. In this representation of a delayed feedback system's dynamics \cite{Giacomelli1996}, the $x$-axis represents the time within one round-trip, which is equivalent to twice the feedback delay time, while the $y$-axis indicates the number of round trips. The diagram shows a near vertical straight banded structure which indicates the robust square oscillations resonant with the chosen comoving reference frame of 12 ns. Fig.~\ref{squareDynamics}d provides an in-depth view of the fast oscillation region, as indicated by the orange square in Fig.~\ref{squareDynamics}c. It is evident that a regular lattice structure is formed by the peaks of the temporal oscillations. The time interval between adjacent peaks is approximately $t_{in} = 0.13$ ns. The stability of this structure indicates a well defined phase relation is preserved from roundtrip to roundtrip. However, the clear slope of this stripe pattern indicates that this fast dynamics is not stationary in the SW reference frame. 

\begin{figure}[ht!]
\centering
  \includegraphics[width=8.5cm]{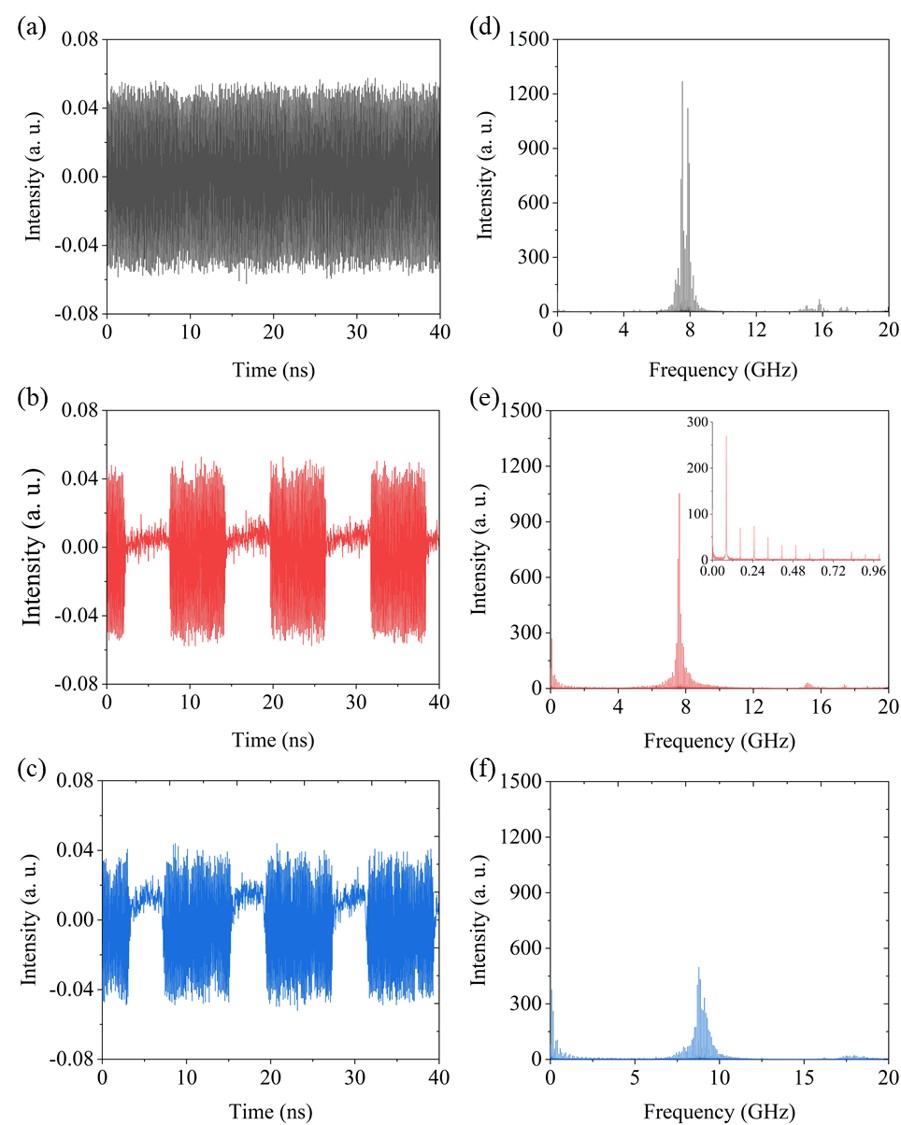}
  \caption{Temporal dynamics and the corresponding RF spectra of the TE mode for different pump currents: (a) and (d), 2.64$P_{th}^{TE}$; (b) and (e), 2.86$P_{th}^{TE}$; (c) and (f) 3.24$P_{th}^{TE}$.}
  \label{Details}
\end{figure}

To further study the formation of the square wave, we investigate the dynamics as a function of the pump current, utilizing a fixed $\lambda/2$ plate set at 96$^\circ$. As shown in Fig.~\ref{Details}, at low pump currents, the laser operates stably, predominantly producing a constant intensity output in the TE mode. The low-level feedback mechanism maintains this stability, ensuring that the laser does not deviate from its operational parameters. As the pump current is increased to 2.64$P_{th}^{TE}$, we observe undamped oscillations in the output with frequency peaks around 7.5 GHz and 7.8 GHz (as depicted in Fig.~\ref{Details}d). The dominant peaks correlate with the observed oscillation frequencies, while weaker modes at 15.6 GHz are also detected. These weaker modes are indicative of higher - order harmonics, which suggest the presence of nonlinearities in the laser's response to the increased pump current.


When the pump current reaches 2.86$P_{th}^{TE}$, distinct and robust square fast oscillations are observed. Those oscillations are characterized by a duration of $\tau_2 = 6.68$ ns and a steady interval of $\tau_1 = 5.40$ ns (illustrated in Fig.~\ref{Details}b). The total period of this temporal structure is still $\tau_1 + \tau_2 = 12.08$ ns, corresponding to twice the delay feedback time. The RF spectrum at this stage shows a dominant peak at 7.6 GHz, with satellite peaks which correspond to the external cavity mode spacing (Fig.~\ref{Details}e). Accordingly, the low frequency region of the power spectrum shows peaks separated by 80~MHz with a strong dominance of the odd harmonics as expected for a square wave signal (as shown in the inset of Fig.~\ref{Details}e). 

Increasing the pump current further to 3.24$P_{th}^{TE}$ results in a notable shift in dynamics. The system now exhibits square oscillations with a diminished intensity, presenting an asymmetric square wave profile (shown in Fig.~\ref{Details}c). Specifically, the duration of the oscillations broadens to $\tau_2 = 8.13$ ns, while the steady interval shortens to $\tau_1 = 3.87$ ns. This shift corresponds with a reduced intensity in the RF spectrum and a central frequency shift to higher values, indicating a change in the relaxation oscillation frequency of the TE mode. Additionally, the prominence of low-frequency harmonics suggests a transition towards chaotic behavior within the laser system.

These observations in the frequency domain indicate that stable, high amplitude fast oscillations within the square wave regions are strongly enhanced near 7.6~GHz and are restricted to a rather narrow current region slightly below 3~mA. In that current region, the relaxation oscillation frequency of the standalone TE mode is very close to 7.6~GHz (see inset of Fig.\ref{lasercurvespectrum}). Thus, we conjecture that these oscillations result from a resonance between the TE relaxation oscillations and the birefringence. Birefringence control is unfortunately not possible in our current experimental setup.

\begin{figure}[ht!]
\centering
  \includegraphics[width=8.5cm]{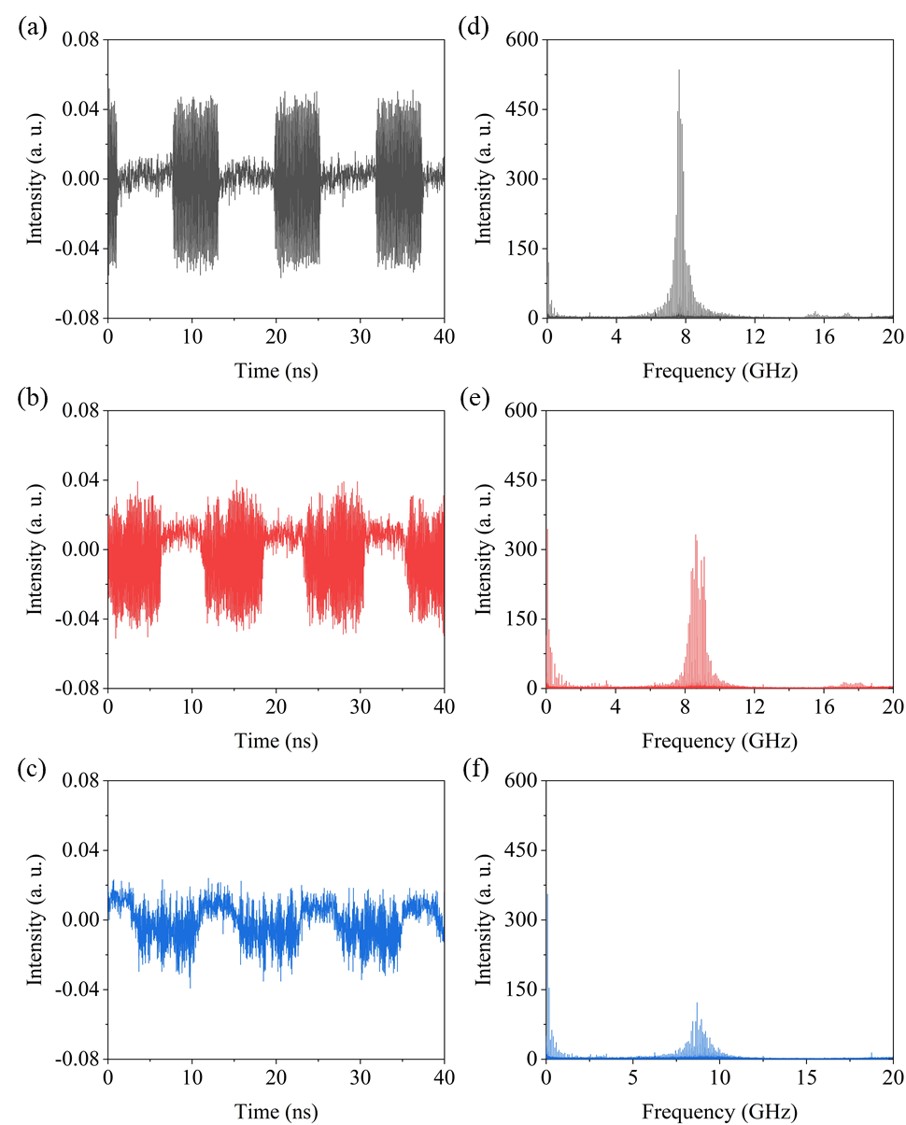}
  \caption{Temporal dynamics and the corresponding RF spectra of the TE mode for different $\theta$: (a) and (d), 90$^\circ$; (b) and (e), 150$^\circ$; (c) and (f) 240$^\circ$.}
  \label{Squarewithplate}
\end{figure}

Subsequently, we delve into the influence of the $\lambda/2$ plate on the temporal dynamics of the laser output. Laser systems often exhibit complex dynamics characterized by the interplay of TE and TM modes. The introduction of optical elements, such as a $\lambda/2$ plate, allows for manipulation of these modes. The $\lambda/2$ plate rotates the polarization of incoming light, effectively altering the phase and amplitude of the TE and TM modes. Therefore, we maintain a constant pump current at 2.87$P_{th}^{TE}$ and vary the angle of the $\lambda/2$ plate. In Fig.~\ref{Squarewithplate}, we can clearly observe square-wave is formed with a duration of $\tau_2 = $5.4 ns when $\theta = 90^\circ$ (Fig.~\ref{Squarewithplate}a). At this angle, the $\lambda/2$ plate effectively rotates the TE and TM modes by 90$^\circ$. The RF spectrum shows the central peak is at 7.6 GHz. Increasing the angle to 150$^\circ$ results in a degradation of the square oscillations. The waveform broadens, indicating a loss of coherence in the TE mode as contributions from the TM mode become significant. The corresponding RF spectrum shows a decrease in intensity, with low-frequency components becoming more pronounced, reflecting the involvement of the TM mode in the dynamics. Upon further rotation of the $\lambda/2$ plate to 240$^\circ$,  the oscillations experience a more severe degradation. The output now manifests as small and wide pulses, with minor oscillations superimposed. The RF spectrum associated with this configuration displays an even lower intensity, indicating a considerable loss of coherence and a more complex interplay between the TE and TM modes.

\begin{figure}[ht!]
\centering
  \includegraphics[width=8.5cm]{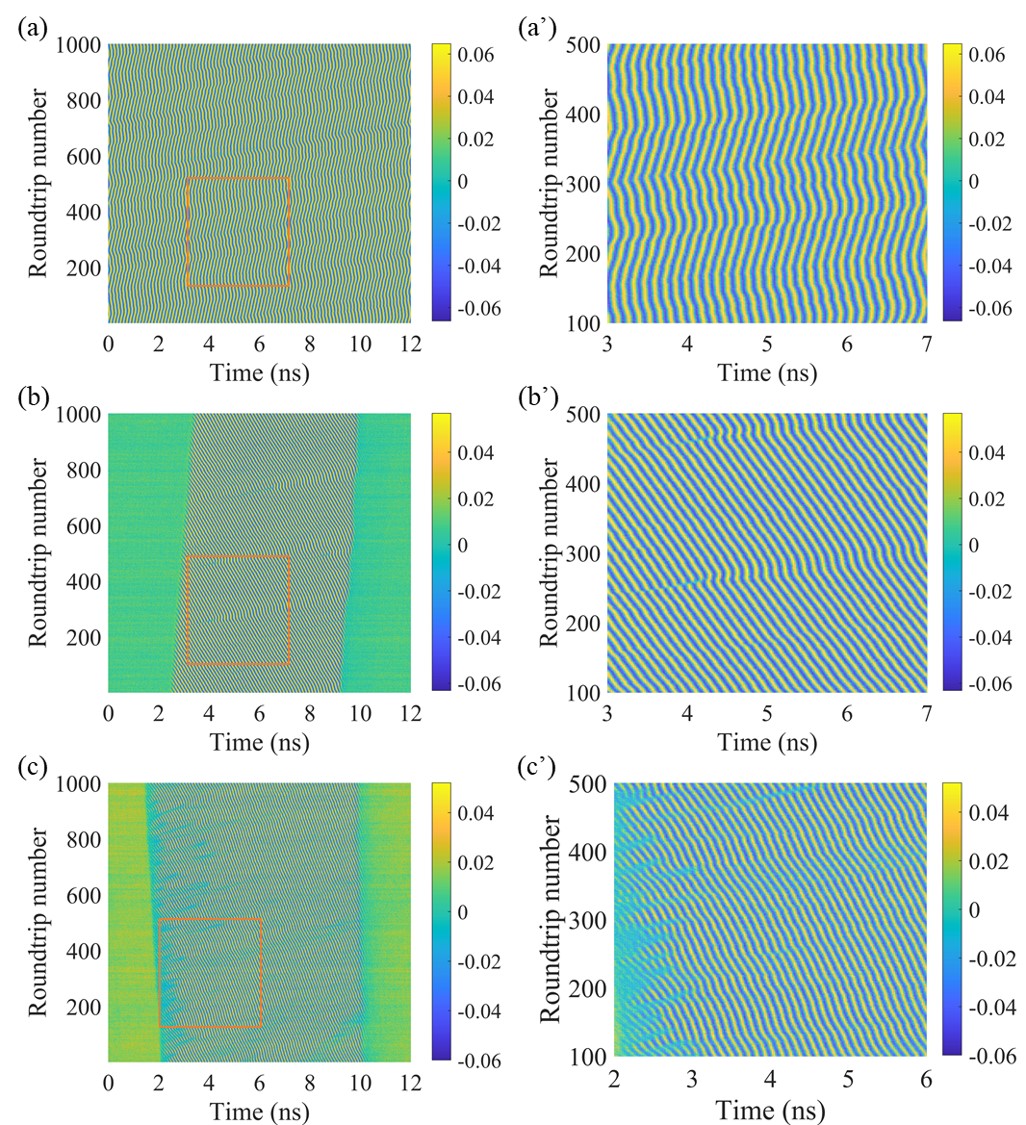}
  \caption{Spatial-temporal dynamics of the TE mode for different pump currents: (a) and (a'), 2.64$P_{th}^{TE}$; (b) and (b'), 2.86$P_{th}^{TE}$; (c) and (c') 3.24$P_{th}^{TE}$. (a')-(c') are the enlarge regions remarked in a-c.}
  \label{Squarewithcurrent}
\end{figure}

Finally, we also study the spatial-temporal dynamics of the TE mode for the laser is operated at different pump currents. Fig.~\ref{Squarewithcurrent}a-c are the dynamics within one round trip as a function of round-trip number, and Fig.~\ref{Squarewithcurrent}a'-c' show the detail structure corresponding to the region within the orange square in Fig.~\ref{Squarewithcurrent}a-c. At a pump current of 2.64$P_{th}^{TE}$, our analysis revealed fast oscillations that produced a distinct zigzag structure in the spatial-temporal domain (see Fig.~\ref{Squarewithcurrent}a and a'). This pattern suggests that the fast oscillation is almost perfectly resonant with the feedback time but that its phase is still wandering in the course of many roundtrips.
Increasing the pump current to 2.86$P_{th}^{TE}$ resulted in a notable transition in the mode structure. A banded structure emerged, extending spatially and indicating the presence of square oscillations (Fig.~\ref{Squarewithcurrent}b). Again, a stable (but non-stationary) pattern emerges, suggesting long-term coherence of the fast dynamics. Upon closer examination (Fig.~\ref{Squarewithcurrent}b'), we observed defects within the pattern, in (4.2, 250) and (4.5, 460). We conjecture that these defects arise from the duration of the modulated phase not being an exact integer number of periods of the fast oscillation. Since the domain enclosing these defects is finite, we can expect their number to evolve in time (see~\cite{arecchi1992patterns}) for a seminal paper in the spatial domain) but a statistical analysis of defect is left for future work. Further increasing the pump current to 3.06$P_{th}^{TE}$ led to the observation of an even broader banded structure (Fig.~\ref{Squarewithcurrent}c). However, this higher level of feedback resulted in the formation of a larger number of defects near the left boundary indicating a degradation of coherence within temporal pattern of the TE mode. The details captured in Fig.~\ref{Squarewithcurrent}c' reveal that these defects, which disrupt the previously observed coherent oscillations. Accordingly, the left and right fronts on Fig.~\ref{Squarewithcurrent}c are not strictly stationary with respect to each other.

\section{Conclusions}
In summary, we have successfully generated fast oscillations of the TE mode over a square wave polarization switching at a frequency different from the relaxation frequency and from the external-cavity frequency in a semiconductor VCSEL. Orthogonal polarization feedback induces a slow SW modulation of the laser polarizations and fast oscillations with possibly high modulation amplitude are observed on the upper state of the SW. Interestingly, although the polarization switching dynamics is induced by delayed optical feedback, that self-pulsation frequency does not scale with the time-delay. Instead, from free-running laser power spectra, we conjecture that this fast part of the dynamics is related to a resonance between the relaxation oscillation frequency of the TE mode and the laser birefringence. Since the whole experiment is performed below the standalone emission threshold of TM, we infer that TM is not driving the instabilities we observe but further polarization-resolved experiments are need to fully address this point. We have investigated the influence of the feedback level induced by the pump current, and the angle of the $\lambda/2$ plate, on the evolution of the dynamics of the TE mode. The results reveal that the resonance between the relaxation oscillation frequency of the TE mode and the longitudinal modes of the external cavity has significant influence on the generation of the square fast oscillations. 
By plotting the spatial-temporal structure, we have identified some defects within the fast oscillation structure which we associate to near-resonance situation between the fast dynamics and the square wave polarization switching. Further analysis of the dynamics of the fronts and statistics of the defects in the spatiotemporal representation is left for future work. 

Our system introduces a streamlined approach for generating tunable square-wave pulses with adjustable durations. By leveraging dynamic front interactions between oscillatory and homogeneous states, the platform enables precise temporal control over pulse characteristics, a feature critical for advanced photonic systems. Notably, the stabilization mechanism—inspired by oscillatory tail-mediated front dynamics (as in Ref.~\cite{Koch2022}—may enable robust formation of temporal localized states, a capability distinct from conventional methods reliant on static or spatially confined structures.

\section*{Acknowledgment}
The authors would like to thank Prof. Gang Xu for the insightful discussions. This work is partially supported by National Natural Science Foundation of China (Grant No. 61804036, and 62475206), Key Research and Development Plan of Shaanxi Province, China (Grant No. 2024GH-ZDXM-42), National Key Research and Development Program of China (Grant No. 2021YFB2801900, 2021YFB2801901, 2021YFB2801902, and 2021YFB2801904). 



\bibliography{biblio}

\end{document}